\documentstyle[aps,epsf,psfig]{revtex}
\begin{document}
\draft

\title {The Self-Trapping Line of the Holstein Molecular Crystal Model in One Dimension}

\author{A.~H.~Romero${^{1,3,}}$\footnote{Present address: Max-Planck Institut f\"{u}r Festk\"{o}rperforschung, Heisenbergstr. 1, 70569 Stuttgart, Germany}, David W. Brown${^2}$ and Katja Lindenberg${^3}$}

\address
{${^1}$
Department of Physics,\\
University of California, San Diego, La Jolla, CA 92093-0354}

\address
{${^2}$
Institute for Nonlinear Science,\\
University of California, San Diego, La Jolla, CA 92093-0402}

\address
{${^3}$
Department of Chemistry and Biochemistry,\\
University of California, San Diego, La Jolla, CA 92093-0340} 

\date{\today} 

\maketitle

\begin{abstract}

The ground state of the Holstein molecular crystal model of a single electron
in one dimension
is studied using the Global-Local variational method, analyzing in particular
the total energy, kinetic energy, phonon energy, and interaction energy
over a broad region of the polaron parameter space.
Through the application of objective criteria, a curve is identified
that simply, accurately, and robustly locates the self-trapping transition
separating small polaron and large polaron behavior.
Particular attention is given to the kinetic energy, which is compared
quantitatively with other perturbative and non-perturbative methods.

\end{abstract}


\narrowtext

\section{Introduction}

Despite the extensive effort and creativity that has been applied to the polaron problem for more than half a century, there remain significant aspects of polaron structure and behavior that have defied satisfying explanation and quantitative description.
Central among these is the so-called self-trapping transition, manifested as a strong feature in the intermediate-coupling regime that separates states that are predominantly ``small-polaron-like'' from states that are predominantly ``large-polaron-like''.

Though this transition is expected to be smooth at finite parameter values, and though perturbation theories exist providing asymptotically accurate descriptions of polaron structure and properties on each side of this transition, the intermediate-coupling region in which the transition lies has proven very resistant to analysis.

Although the physically-meaningful self-trapping transition is not a singular feature of the polaron landscape at finite parameter values, the self-trapping transition is related to singular behavior in the adiabatic limit.
This singular behavior in the far reaches of parameter space appears to influence both weak-coupling perturbation theory and strong coupling perturbation theory at finite parameter values in that weak-coupling perturbation theory
\cite{Migdal58,Nakajima80,Mahan93,Alexandrov95,Romero98d,Romero98g,Romero99}
breaks down as this feature is approached from below, and strong-coupling perturbation theory
\cite{Romero98g,Romero99,Tyablikov52,Lang63,Gogolin82,Marsiglio95,Stephan96,Capone97}
breaks down as this feature is approached from above.

Much as a discrete curve (e.g., the unit circle) may describe a limited domain of convergence for a series representation of a function though the function itself may be well-behaved across most of that curve, there is meaning to the notion of a discrete line describing the location of the self-trapping transition though the physical phenomenon we study may be smooth throughout the finite parameter space.
However, lacking absolute knowledge of the underlying self-trapping phenomenon, we must necessarily proceed more empirically and attempt to {\it infer} the location of a broadly-meaningful self-trapping line from limited observations of polaron properties.
From such observations on numerous distinct quantities over large regions of the polaron parameter space, we conclude that the self-trapping line of the Holstein model in one dimension is accurately and robustly described by the simple relation
\begin{equation}
g_{ST} = 1 + \sqrt{J/\hbar \omega} ~,
\label{eq:gst}
\end{equation}
in which $J$ is the nearest-neighbor electronic hopping integral, $\omega$ is the Einstein frequency, and $g$ is the dimensionless coupling strength.
By this we do {\it not} mean that every physical property exhibits transition behavior on this curve at all parameter values, but that {\it collectively} transition behaviors in all measured properties are consistently and quantitatively related to this single curve.
This curve could already be inferred roughly in Figure 4 of Ref. \cite{Brown97b}, but here we are led to this functional form by direct, high-precision numerical study.

Describing the self-trapping transition with accuracy clearly requires methods that are {\it non-perturbative}.
In recent years, several methods have been developed that are capable of describing the intermediate-coupling regime with high accuracy over non-trivial ranges of adiabaticity;
these include
variational techniques ~\cite{Romero98g,Brown97b,Brown97a,Zhao97a,Zhao97b,Romero98a,Romero98e,Trugman99,Bonca98},
cluster diagonalization ~\cite{Capone97,Wellein97a,deMello97,Alexandrov94a},
density matrix renormalization group 
(DMRG) ~\cite{Jeckelmann98a,Jeckelmann98b} and
quantum Monte Carlo simulations (QMC) ~\cite{McKenzie96,DeRaedt83,DeRaedt84,Lagendijk85,Kornilovitch98a,Alexandrov98a,Kornilovitch99}.

Elsewhere \cite{Romero98a}, we have made direct quantitative comparisons among a number of these and other independent methods, demonstrating broad quantitative agreement among the best.
Here, we use the Global-Local variational method on periodic lattices of 32 sites, supported by low orders of perturbation theory on infinite lattices.
Details about the numerical method can be found in Refs.~\cite{Romero98g,Brown97b}.

We focus on the 1-D Holstein Hamiltonian for a single electron ~\cite{Holstein59a,Holstein59b}

\begin{equation}
\hat{H} = \hat{H}_{kin} + \hat{H}_{ph} + \hat{H}_{int} ~,
\end{equation}
\begin{equation}
\hat{H}_{kin} = - J \sum_n a_n^{\dagger} ( a_{n+1} + a_{n-1} ) ~,
\end{equation}
\begin{equation}
\hat{H}_{ph} = \hbar \omega \sum_n b_n^{\dagger} b_n ~,
\end{equation}
\begin{equation}
\hat{H}_{int} = - g \hbar \omega \sum_n a_n^{\dagger} 
a_n ( b_n^{\dagger} + b_n ) ~,
\end{equation}
in which $a_n^\dagger$ creates an electron in the rigid-lattice Wannier state at site $n$, and $b_n^\dagger$ creates a quantum $\hbar \omega$ of vibrational energy  in the Einstein oscillator at site $n$; $J$ is the electronic transfer integral between nearest-neighbor sites, and $g$ is the local electron-phonon coupling constant. 

\section{Characteristic Energies}

The global ground state energy $E_0$ is composed of contributions from the three principal components of the total Hamiltonian:
\begin{equation}
E_0 = \langle \psi | H | \psi \rangle 
= E_{kin} + E_{ph} + E_{int}  ~,
\end{equation}
in which $E_{kin}$ is the kinetic energy, $E_{ph}$ is the phonon energy, and $E_{int}$ is the electron-phonon interaction energy.

\begin{figure*}[t]
\begin{center}
\leavevmode
\epsfxsize = 3.5in
\epsffile{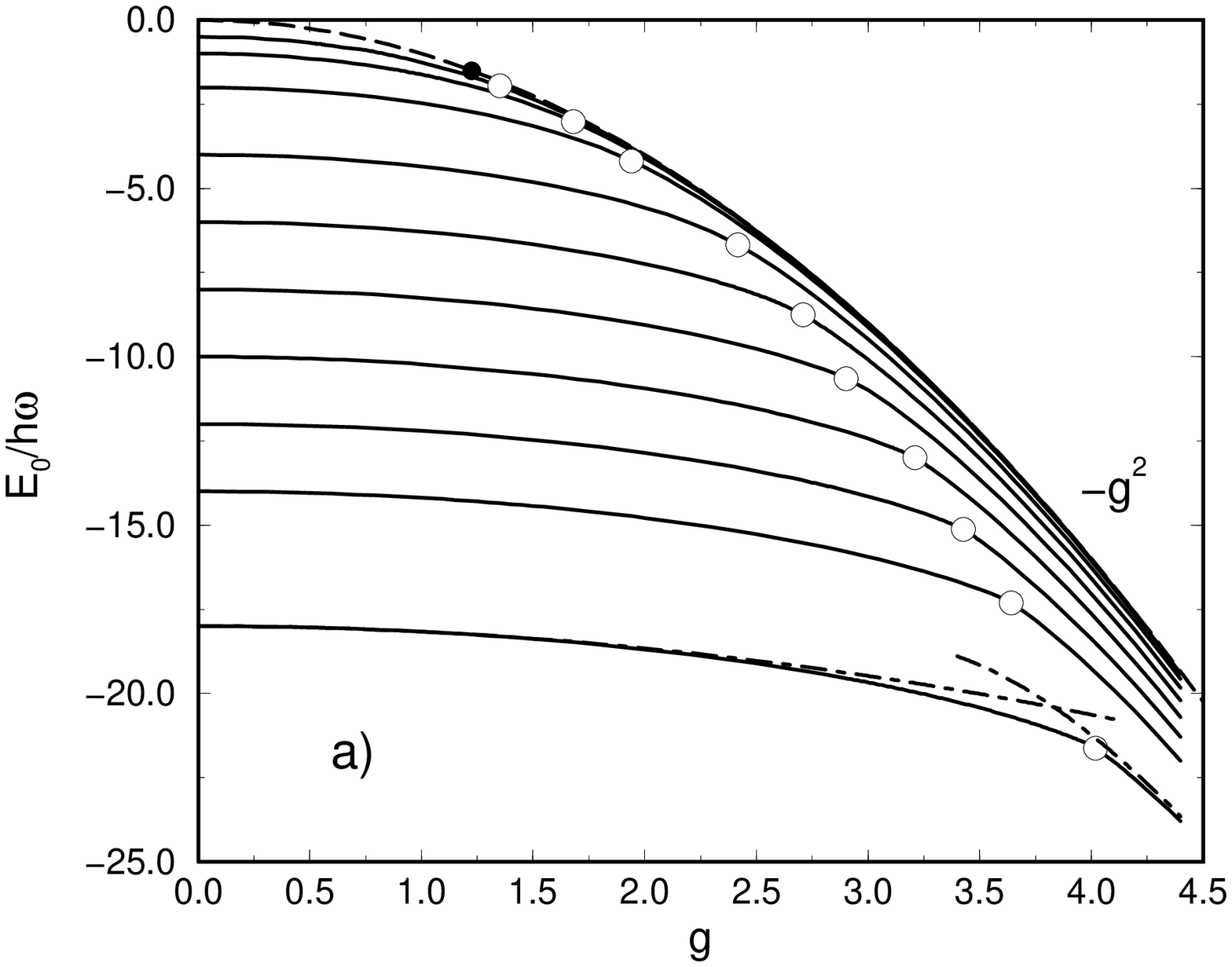}
\vspace{.05in}
\leavevmode
\epsfxsize = 3.5in
\epsffile{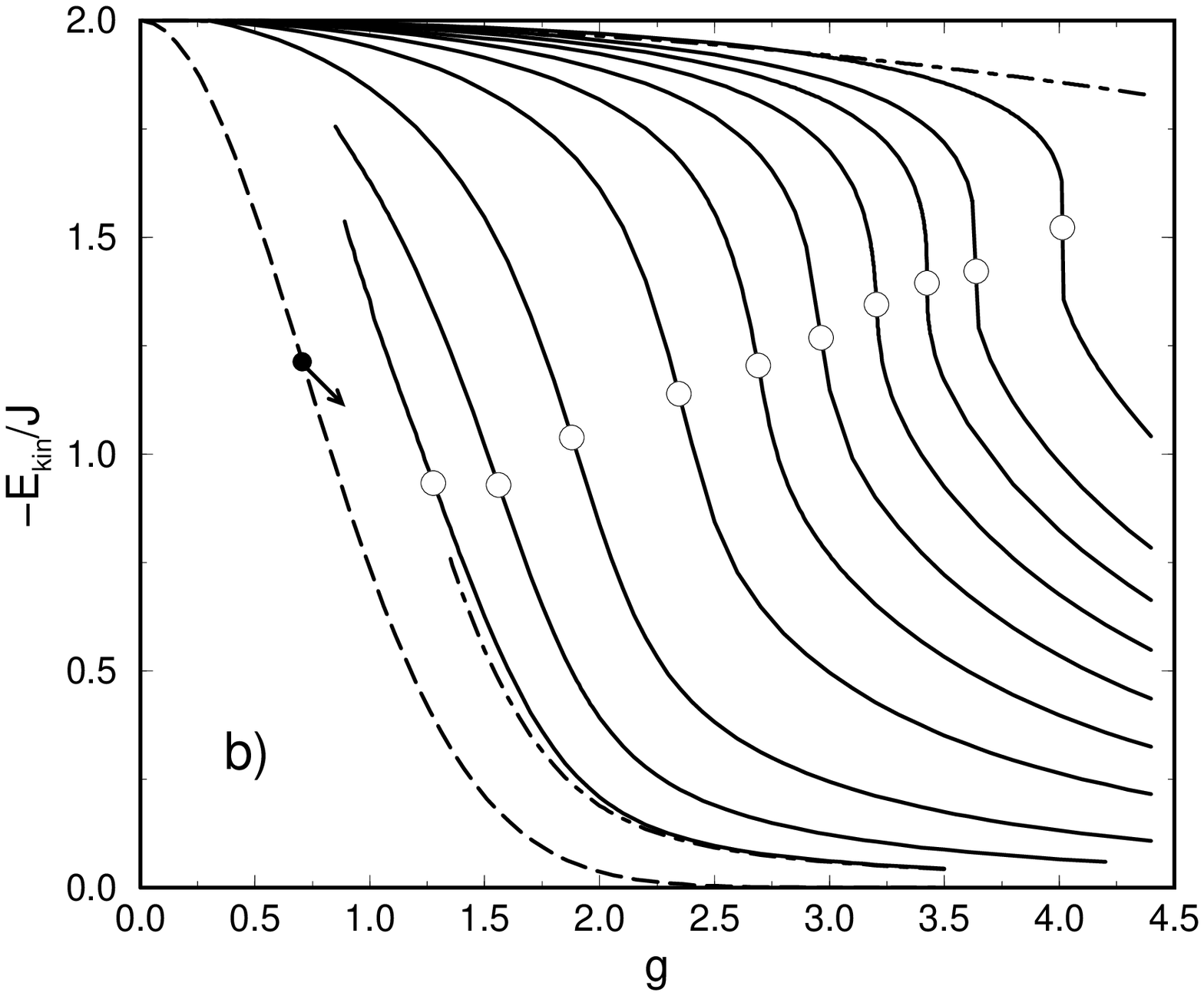}
\end{center}
\begin{center}
\leavevmode
\epsfxsize = 3.5in
\epsffile{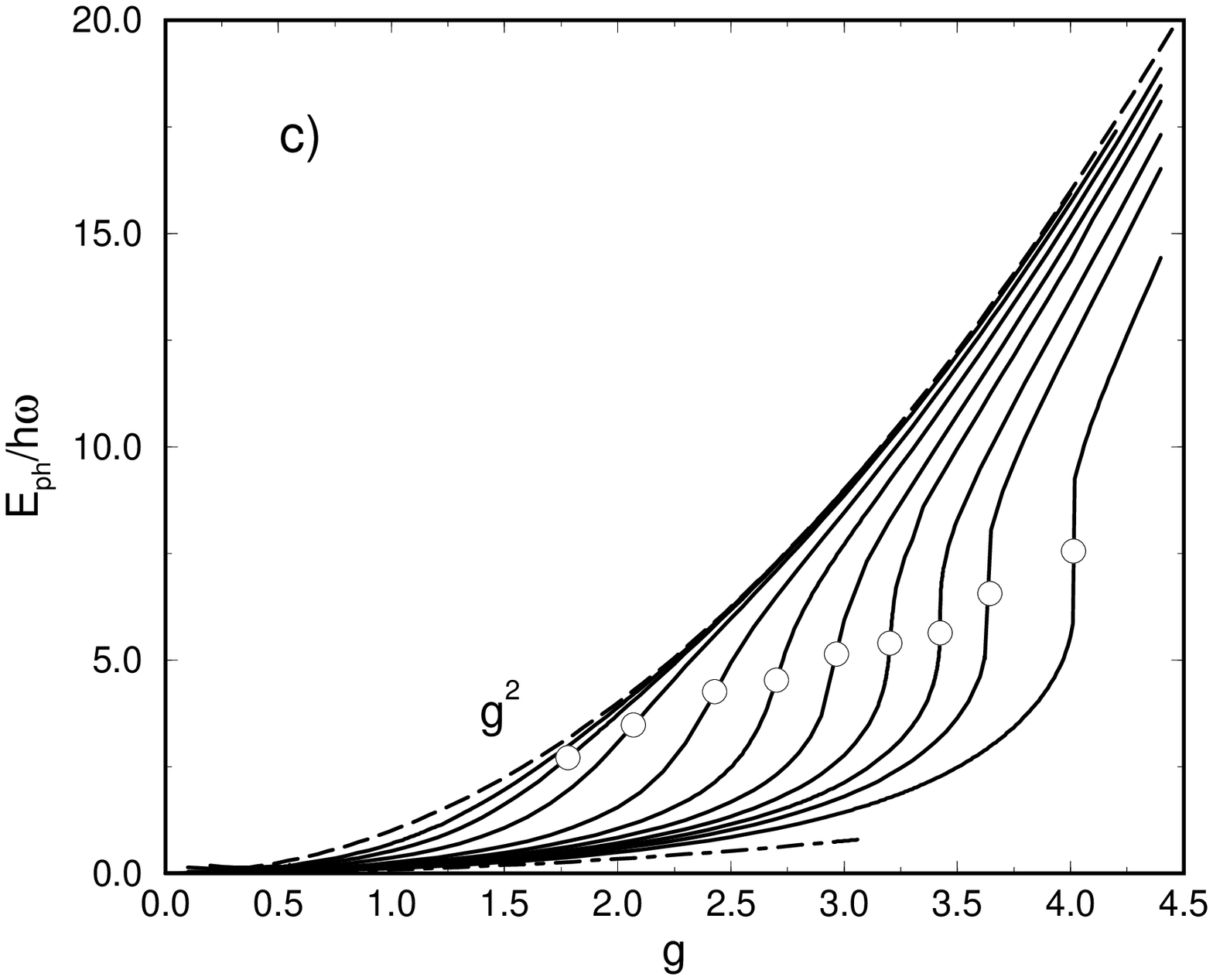}
\vspace{.051in}
\leavevmode
\epsfxsize = 3.5in
\epsffile{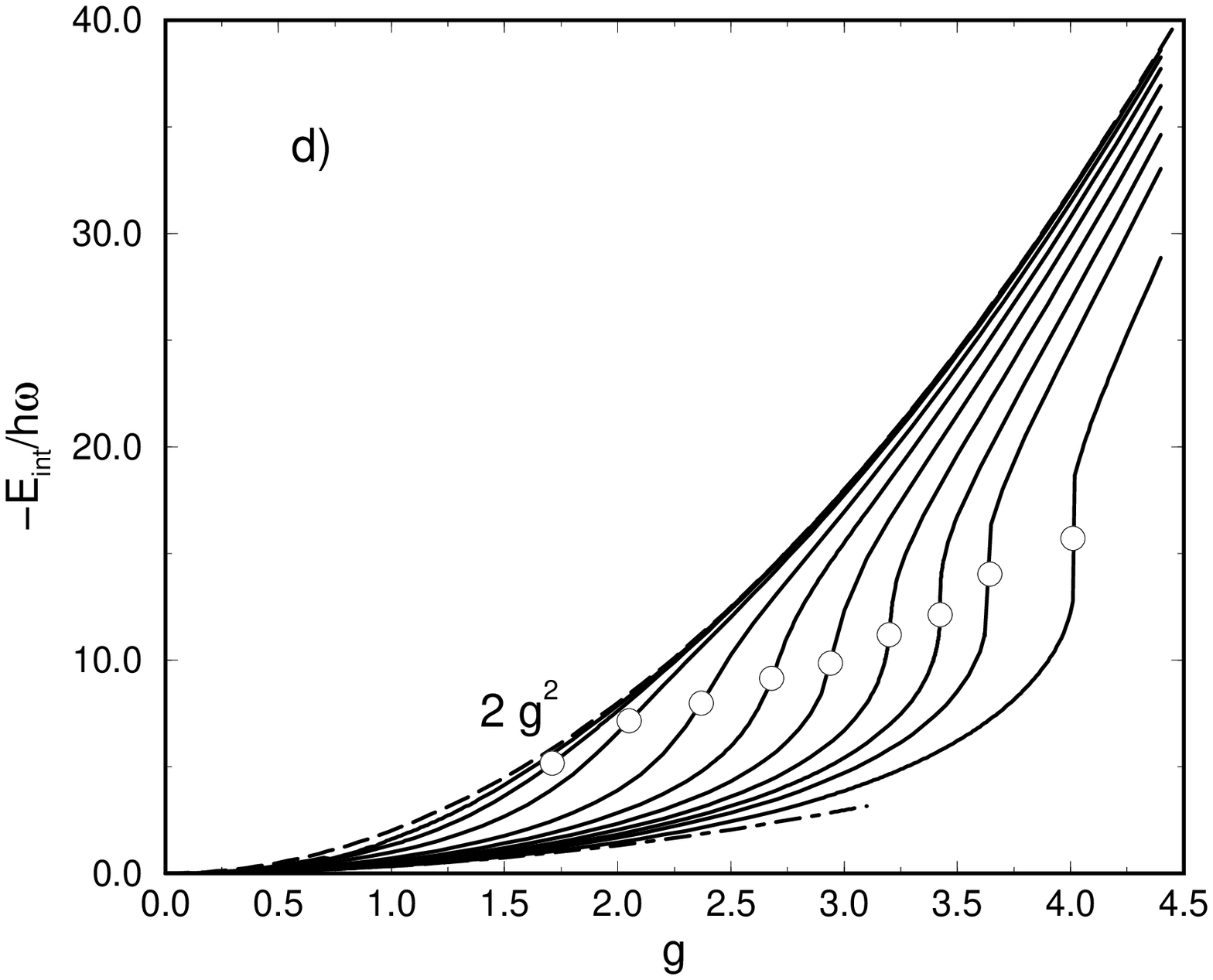}
\end{center}
\caption
{
The ground state energy $E_0$ (a) and its several components $E_{kin}$ (b), $E_{ph}$ (c), and $E_{int}$ (d) as functions of the electron-phonon coupling for $J/ \hbar \omega = 0.25$, $0.5$, $1.0$, $2.0$, $3.0$, $4.0$, $5.0$, $6.0$, $7.0$ and $9.0$ (solid curves, top to bottom in (a), left to right in (b), (c), and (d)). 
The dashed curve in each figure panel is the exact $J \rightarrow 0$ limit appropriate to each case.
The chain-dotted curves illustrate Eqs.~(\ref{eq:E0WC})~-~(\ref{eq:E0SC}) for $J/\hbar \omega = 9$, and (\ref{eq:EkinSC}) for $J/ \hbar \omega = 1/4$.
Circles ($\circ$) indicate the ``break'' in each energy curve associated with the self-trapping transition.
Bullets ($\bullet$) in (a) and (b) indicate the $J = 0$ termini of the self-trapping lines appropriate to $E_0$ and $E_{kin}$ as determined from Eqs. (\ref{eq:E0SC}) and (\ref{eq:EkinSC}), respectively.
The arrow in (b) indicates the initial slope of the $E_{kin}$ self-trapping line as determined from Eq. (\ref{eq:EkinSC}).
}
\label{fig:energies}
\end{figure*}

Our numerical results for these quantities are summarized in Figure~\ref{fig:energies} as functions of the electron-phonon coupling strength $g$ for several values of $J/ \hbar \omega$.
Each curve in Figure~\ref{fig:energies} contains 80-200 data points and each figure panel collectively represents nearly 1200 distinct polaron ground states.
No smoothing has been performed; each ``curve'' is a polygonal arc connecting computed energies.

Over most of the parameter space we have investigated, computational errors are smaller than can be meaningfully conveyed with any resolvable symbol; the principal exception is at weak coupling and smaller $J$ values where decreasing sensitivity of the ground state energy to certain details of polaron structure eventually hampers convergence.

With increasing adiabaticity, here beginning at $J/\hbar \omega \approx 7$, the ability of the variational method to represent the complexity of polaron structure in the immediate vicinity of the self-trapping transition eventually is overtaxed, and discontinuities appear in estimated quantities such as the kinetic energy.
Although the {\it value} of the kinetic energy in the immediate vicinity of such anomalies is necessarily distorted, the location of the discontinuities continues to provide reasonable estimates for the location of the self-trapping transition, and outside of a narrow region, quantitative accuracy remains good \cite{Romero98a}.
For such reasons, in the following we retain data for the cases $J/\hbar \omega = 7$ and $9$.

\widetext
\narrowtext

Although the aggregate ($E_0$) of these several contributions is subject to variational constraint, the values of the separate contributions ($E_{kin}$, $E_{ph}$, $E_{ph}$) are not; in principle, the latter are vulnerable to distortions that may misrepresent some aspects of ground state structure while still yielding favorable results for the ground state energy.
It is for such reasons that in the course of developing our method and in using it to obtain new results, we pay close attention to the detail of overall polaron structure and compare multiple quantities with known results available from independent approaches.

In the appropriate regimes, the energy contributions shown in Figure~\ref{fig:energies} are in excellent agreement with the weak-coupling perturbation theory results (Rayleigh-Schr\"{o}dinger, $\hat{H}_0 = \hat{H}_{kin} + \hat{H}_{ph}$, $\hat{H}' = \hat{H}_{int}$) \cite{Nakajima80,Mahan93,Alexandrov95,Romero98g}
\begin{eqnarray}
E_0^{WC} & \sim & - 2J - g^2 \frac {\hbar \omega} {\sqrt{1+4J/ \hbar \omega}} ~,
\label{eq:E0WC} \\
E_{kin}^{WC} & \sim & - 2 J + g^2 \frac {2J} { (1+4J/ \hbar \omega)^{3/2} } ~,
\label{eq:EkinWC} \\
E_{ph}^{WC} & \sim & g^2 \left[ \frac {\hbar \omega} {\sqrt{1+4J/ \hbar \omega}} - \frac {2J} { (1+4J/ \hbar \omega)^{3/2} } \right] ~, \\
E_{int}^{WC} & \sim & - g^2  \frac {2 \hbar \omega} {\sqrt{1+4J/ \hbar \omega}} ~,
\end{eqnarray}
and with the strong-coupling perturbation theory results (Rayleigh-Schr\"{o}dinger following the Lang-Firsov transformation, $\hat{H}_0 = \tilde{H}_{ph} + \tilde{H}_{int}$, $\hat{H}' = \tilde{H}_{kin}$) \cite{Alexandrov95,Romero98g,Lang63,Marsiglio95,Stephan96}
\begin{eqnarray}
E_0^{SC} &\sim& -g^2 \hbar \omega \nonumber \\
&& - 2 J e^{-g^2} \nonumber \\
&& - \frac {2 J^2} {\hbar \omega} e^{ -2 g^2} [ f(2 g^2) + f(g^2) ]  ~,
\label{eq:E0SC} \\
&\sim& -g^2 - \frac {J^2} {g^2} ~~~~~ ~~~~~ g \gg 1 \\
E_{kin}^{SC} &\sim& -2 J e^{-g^2} \nonumber \\
&& - \frac {4 J^2} {\hbar \omega} e^{ -2 g^2} [ f(2 g^2) + f(g^2) ] ~,
\label{eq:EkinSC} \\
&\sim& - \frac {2 J^2} {g^2} ~~~~~ ~~~~~ ~~~~~ g \gg 1 \\
f \rm (y) &=& \rm{Ei}(y) - \gamma - \ln(y) ~,
\end{eqnarray}
where $\gamma$ is the Euler constant and $\rm{Ei}(y)$ is the exponential
integral.
One sample curve representing each of Eqs. (\ref{eq:E0WC}) - (\ref{eq:EkinSC}) has been included in Figure~\ref{fig:energies} to illustrate this agreement.
More detailed discussion and comparisons with other theories have been given elsewhere~\cite{Romero98a}.

The kinetic energy is particularly important as a ground state property that is intimately connected with electron mobility.
Characteristically, the kinetic energy is a weak function of the electron-phonon coupling below the self-trapping transition, and this dependence grows increasingly weak with increasing adiabaticity.
Owing to the minimal involvement of phonons in the polaron in this regime, the quasi-particle can be fairly characterized as a quasi-free electron with a slightly reduced bandwidth.
At very strong coupling, the kinetic energy decays to zero, suggesting that the dressed electron becomes essentially immobile relative to the quasi-free electron; significant, however, is the fact that this decay is not ultimately exponential in the coupling constant as is commonly assumed on the basis of the small polaron approximation (first order SCPT), but a much weaker inverse power as suggested by perturbative corrections (second-order SCPT) \cite{Alexandrov98}.

\begin{figure*}[htb]
\begin{center}
\leavevmode
\epsfxsize = 3.5in
\epsffile{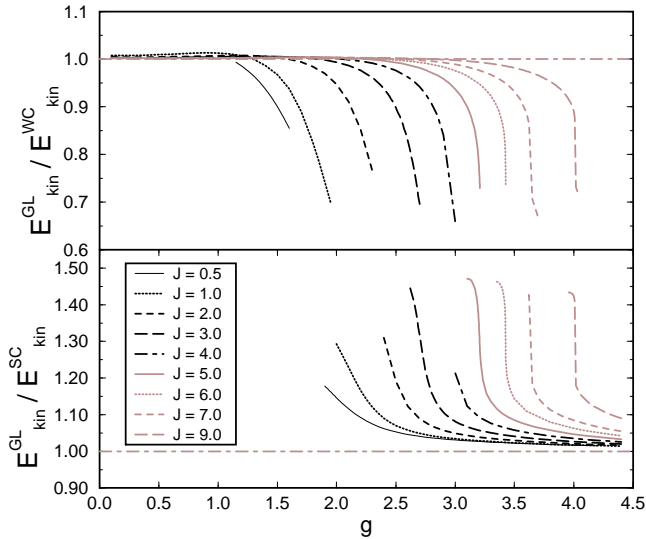}
\end{center}
\caption
{
Comparison of Global-Local kinetic energy, $E_{kin}^{GL}$, with weak-coupling perturbation theory, $E_{kin}^{WC}$ according to (\ref{eq:EkinWC}) (upper panel), and with the Lang-Firsov approximation, $E_{kin}^{SC}$ according to (\ref{eq:EkinSC}) (lower panel), as functions of the electron-phonon coupling. 
}
\label{fig:ekinvspert}
\end{figure*}

A comparison of our numerical kinetic energies with weak (\ref{eq:EkinWC}) and strong (\ref{eq:EkinSC}) coupling perturbation theories is shown in Figure~\ref{fig:ekinvspert}.
Each of these comparisons exhibits sharp deviations in the vicinity of the self-trapping transition because neither WCPT nor SCPT undergoes the transition while the variational results do.
Apart from data affected by deteriorating precision, the agreement between our variational and perturbative kinetic energy in the weak-coupling regime is very good up to $g$ values close to the self-trapping transition.
For this weak coupling case, perturbation theory characteristically {\it over}estimates the kinetic energy value, {\it under}representing the integration of phonons into polaron structure, causing the electron to appear more ``free'' than it actually is.
This is to be expected, of course, since the inherent anharmonicity of the large polaron is induced by the electron-phonon interaction even at weak coupling, but is only partially captured by the low orders of weak-coupling perturbation theory.

For strong coupling, our variational calculation approaches the second-order SCPT result above the self-trapping transition; however, deviations persist significantly into the strong coupling regime, with larger $J$'s converging to the SCPT result more slowly than smaller $J$'s.
Strong coupling perturbation theory systematically {\it under}estimates the kinetic energy, {\it over}representing the integration of phonons into polaron structure, causing the electron to appear more ``trapped'' than it actually is.
This, too, is to be expected, since the zeroth-order state of strong coupling perturbation theory is the extreme limit of a completely localized excitation; the finite real-space spread of the true self-consistent polaron state is only partially captured at low orders.

\begin{figure*}[htb]
\begin{center}
\leavevmode
\epsfxsize = 3.75in
\epsffile{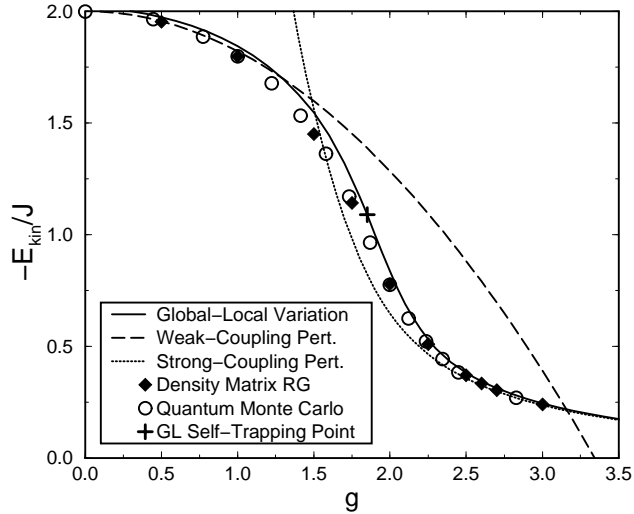}
\end{center}
\caption
{
Comparison of the kinetic energy $E_{kin}$ as determined by the Global-Local method (solid line) with weak coupling perturbation theory (dashed line), strong coupling perturbation theory (dotted line), quantum Monte Carlo simulation (scatter plot), and density matrix renormalization group (diamonds) for $J/\hbar \omega =1$.
The (+) symbol indicates the location of the self-trapping transition as determined here.
QMC data kindly provided by P. E. Kornilovitch \protect \cite{Kornilovitch99} \protect.
DMRG data kindly provided by E. Jeckelmann \protect \cite{Jeckelmann98b} \protect.
}
\label{fig:ekinvsqmc}
\end{figure*}

While this mutual consistency among limiting results is satisfying, the quantitative accuracy of both weak- and strong-coupling perturbation theory is superceded by that of the Global-Local method at intermediate coupling where the self-trapping transition is found.
Figure~\ref{fig:ekinvsqmc} illustrates this point by comparing WCPT and SCPT with the Global-Local variational result as well as with results from quantum Monte Carlo simulation and the density matrix renormalization group approach.
In the following, we rely exclusively upon our variational method for locating the transition.

\section{Self-Trapping Line}

The self-trapping transition is the more-or-less rapid change in polaron structure from that typical of large polarons (below the transition) to that typical of small polarons (above) as $g$ or $J$ are varied, and is typically evident in features that grow increasingly sharp with increasing adiabaticity.
The self-trapping transition is certainly the most exotic feature of the polaron phase diagram, and is intimately involved with, if not always ultimately responsible for, many of the difficulties encountered in polaron theory.
The {\it physical} transition is smooth at finite $g$ and $J/\hbar \omega$ \cite{Lowen88}; however, as exemplified in the previous section, it is common for approximate descriptions of the phenomenon either to miss the transition completely, or to break down in some respect in its vicinity.

In view of the smoothness of the physical self-trapping transition, however, there is no reason to expect the transition to be manifested in exactly the same way in distinct physical quantities; thus, we expect an intrinsic ambiguity in the precise location of the the self-trapping transition that is diminished only by the progressive sharpening of the underlying physical phenomenon.

\begin{figure*}[t]
\begin{center}
\leavevmode
\epsfxsize = 3.6in
\epsffile{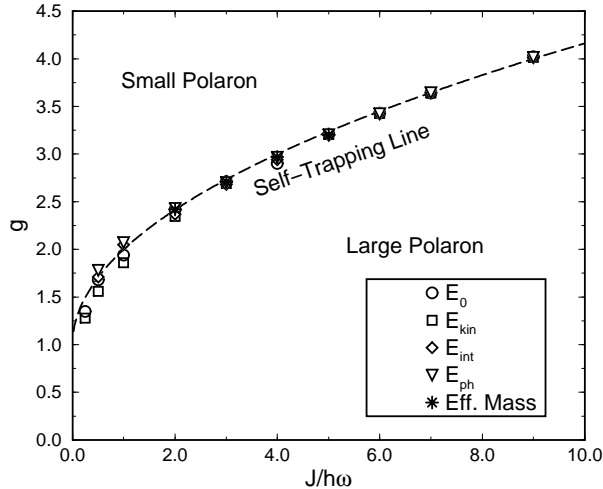}
\end{center}
\caption
{
Polaron phase diagram, showing the location of the self trapping transition.
Discrete symbols correspond to our numerical determinations by applying rapidity criteria to assorted Global-Local energy band data; we have included results from a similar analysis of the polaron effective mass in Ref.
\protect \cite{Romero98e} \protect.
The dashed line corresponds to the empirical curve $g_{ST} = 1 + \sqrt{J/ \hbar \omega}$; no fit has been performed to obtain this curve.
}
\label{fig:phasediagram}
\end{figure*}

Since each of the energies displayed in Figure~\ref{fig:energies} exhibits a ``knee'' or ``break'' that clearly separates distinct weak- and strong-coupling trends, any one of $E_0$, $E_{kin}$, $E_{ph}$, or $E_{int}$ could be used to locate the self-trapping effect.  
As objective criteria for locating this transition, we associate the transition with the particular coupling strength in the intermediate regime where each energy changes {\it most rapidly} with respect to $g$ at fixed $J/ \hbar \omega$; these points are identified by zeros in the second derivatives with respect to $g$ of $E_{kin}$, $E_{ph}$, and $E_{int}$, and in the third derivative of $E_0$ as collected in Figure~\ref{fig:phasediagram}; other rapidity criteria will produce slightly different traces.

These criteria amplify numerical errors and therefore are quite demanding of numerical precision.
Over the interval where the transition was sought, we used a sampling interval $\Delta g$ sufficiently small that directly-computed $g$-values alone would suffice to convey the message of Figure~\ref{fig:phasediagram}.
However, as our convergence criterion, we required sufficient smoothness in the computed derivatives that the determining zeros could be interpolated with absolute numerical error considerably smaller than the sampling interval in most cases.
While this degree of precision is not really necessary to convincingly locate the self-trapping line, we applied to our study the further ``check'' that the collection of numerically-determined self-trapping points should describe sensible curves through the actual family of energies as presented in Figure~\ref{fig:energies}.
This proves to be a stringent error criterion, since the numerical errors already amplified by repeated differentiation with respect to $g$ are further amplified by the characteristic steepness of $E_{kin}$, $E_{ph}$, and $E_{int}$ in the vicinity of the transition.
It is especially significant, therefore, that the numerically-determined self-trapping loci obtained from $E_{kin}$ not only describe a ``sensible'' curve, but appear to connect smoothly with the ascertainable trend as $J/\hbar \omega \rightarrow 0$ (See Figure~\ref{fig:energies}b).

Such considerations permitted a clear resolution of the self-trapping transition for most hopping integral values used in this work, with a characteristic deterioration of precision at small $J/ \hbar \omega$ and $g$ here affecting primarily $J/ \hbar \omega < 1$ and $g< 1.5$.
For this reason, no self-trapping loci are reported for $E_{ph}$ and $E_{int}$ at $J/\hbar \omega = 1/4$.

Though precision deteriorates at small $J/ \hbar \omega$, the dispersion among estimated self-trapping locations that is evident in Figure~\ref{fig:phasediagram} at smaller $J/ \hbar \omega$ values is {\it not} the result of numerical errors, but reflects the intrinsic ambiguity in the assignment of a {\it sharp} location to a {\it smooth} transition.
Using perturbation theory (See (\ref{eq:E0SC}) and (\ref{eq:EkinSC})), for example, one can show that as $J/ \hbar \omega \rightarrow 0$, the trend in the kinetic energy criterion leads to $g= \frac 1 {\sqrt{2}}$ while the trend in the ground state energy criterion leads to $g = \sqrt{\frac 3 2}$.
This is consistent with the relative trends in the self-trapping estimates derived from $E_{kin}$ and $E_0$ data, and is at least roughly consistent with the absolute trends.
This intrinsic non-coincidence implies that there is no single line consistent with {\it all} of the criteria that might legitimately be considered to locate the transition.
Thus, the function (\ref{eq:gst}) appears to describe the {\it common} or {\it criterion-independent} trend line about which distinct locators are tightly clustered with a spread that narrows steadily in both relative and absolute terms with increasing adiabaticity.

There is reason to expect (\ref{eq:gst}) to continue to accurately locate the self-trapping transition {\it beyond} the investigated regime, since it is known from strong-coupling theory that the transition is associated with a particular value of the effective coupling parameter $\lambda \equiv g^2\hbar\omega /2J$ in the adiabatic limit \cite{Alexandrov95}
\begin{equation}
\lambda_{ST} \equiv \frac {g_{ST}^2 \hbar \omega} {2J} \rightarrow \lambda_c = \frac 1 2
\label{eq:lambdac}
\end{equation}
which is consistent with (\ref{eq:gst}).

Our conclusion that the relation (\ref{eq:gst}) accurately and robustly describes the location of the self-trapping transition in the Holstein model is borne out as well in the behavior of other physical quantities not discussed here.
Particularly important among these is the polaron effective mass, which has been discussed in depth in Ref.~\cite{Romero98e}, whose consistency with our present results is illustrated in Figure~\ref{fig:phasediagram}.
The effective mass is of great importance because it is a widely-recognized hallmark of the self-trapping transition and because it demonstrates that the scope of the self-trapping transition extends beyond the ground state.
Other measures at the heart of the polaron problem include the electron-phonon correlations that define the spatial structure of the polaron; presumably the self-trapping transition should involve characteristic rapid changes in spatial structure (e.g., ``localization'' upon self-trapping) that can be analyzed in order to assess the location of the transition.
An extensive analysis of such correlations to be presented elsewhere \cite{Romero98g,Romero99} fully supports the conclusions of this paper in quantitative detail.

\section{On Dimensionality and Adiabaticity}

The results of this paper apply to polarons in one space dimension.
Extrapolating from these results to two and three dimensions is necessarily speculative; however, several pertinent observations can be made.

The results that have long characterized commonly-held expectations for the dimensionality dependence of polaron structure are due to behavior ascertainable in the adiabatic approximation \cite{Sumi73,Emin76,Toyozawa80a}.
Without recounting well-known arguments in detail, we note that these expectations relate to the existence, possible co-existence, and relative stability of infinite-radius and finite-radius states; the former are often referred to as ``free'' states and the latter as ``self-trapped'' states.
``Free'' states are inessentially distinct from weakly-scattered free electron states while ``self-trapped'' states differ markedly from free-electron states in many respects.

In two and three dimensions, the minimum energy states in the adiabatic approximation are found to be ``free'' states throughout the weak-coupling regime up to a discrete (structure-dependent) coupling threshold beyond which ``self-trapped'' states have the minimum energy.
This transition phenomenon is what is meant by the term ``self-trapping transition'' in the adiabatic approximation.
Accordingly, there is no occasion to distinguish large from small polarons in two and three dimensions since the ``free'' states below the transition are of infinite radius and {\it distinct} from large polarons, and the ``self-trapped'' states above the transition are always interpretable as small polarons.
This set of circumstances in two and three dimensions is reflected in the catch phrase ``all polarons are small'', since in this view large polarons in the adiabatic sense are never characteristic of the polaron ground state.

In one dimension, on the other hand, ``free'' states are not found at all in the adiabatic approximation except in the limit of vanishing electron-phonon coupling; instead, finite-radius (i.e. ``self-trapped'') states are found at all finite coupling strengths, leading to the commonly encountered view that there is no self-trapping transition in one dimension.
That these universally ``self-trapped'' states in one dimension might be distinguishable as large polarons or small polarons is of little consequence in this view, and one is led to consider the notion of a resolvable transition between distinct large and small polaron structures as inconsequential as well.
The issue cuts deeper, however, in that in the adiabatic approximation no such characteristic transition from large-to-small polaron structure is found.

That we here find clear and essential transition behavior between large and small polaron states in the one-dimensional case stands in stark contrast to the conventional adiabatic perspective; there is no casual reinterpretation of one or the other set of results that relieves the contrast between such distinct alternatives.
Since our one-dimensional results are quantitatively supported by independent high-quality methods (including cluster diagonalization \cite{Capone97,Wellein97a,deMello97,Alexandrov94a}, density matrix renormalization group \cite{Jeckelmann98a,Jeckelmann98b}, and quantum Monte Carlo \cite{DeRaedt83,DeRaedt84,Lagendijk85,Kornilovitch98a,Alexandrov98a,Kornilovitch99}) we are led to conclude that the problem to be resolved lies not with the data or methodology upon which we base our analysis, but with the adiabatic approximation itself.
Indeed, we need not speculate on this point, since elaborations of adiabatic theory incorporating {\it non-adiabatic} corrections \cite{Alexandrov94a,Kabanov93,Kabanov94} support our overall conclusion that the adiabatic approximation as it is widely regarded fails to embrace non-adiabatic characteristics that are essential to the proper description of polaron states in the weak coupling regime, and therefore fails as well to properly describe the self-trapping transition itself.
This conclusion is deeply rooted in the quantum-mechanical nature of electron-phonon correlations and is not unique to one dimension, suggesting that the adiabatic picture of self-trapping in two and three dimensions may misrepresent the true nature of polaron structure and self-trapping as well.
We need not speculate too broadly on this point either, since an increasing body of results (especially quantum Monte Carlo) consistently point to the {\it absence} of any dramatic qualitative differences between the occurrence of self-trapping in one dimension and in higher dimensions.

The consistent conclusion from all these studies is that in any dimension self-trapping can be understood as a transition between large and small polarons, and that the large polaron state is non-trivially distinct from {\it both} the free electron state {\it and} the large polaron state commonly found in the adiabatic approximation.

It may be suggested that although adiabatic corrections might be expected to be significant at sufficiently small $J/\hbar\omega$, the conventional adiabatic perspective on self-trapping ought to emerge as the adiabatic limit ($J/\hbar\omega \rightarrow \infty$) is properly taken.
As discussed at (\ref{eq:lambdac}) {\it ff.}, however, many arguments going beyond the adiabatic approximation point to a critical self-trapping transition in the vicinity of $\lambda_c [1] = 1/2$ in the adiabatic limit in one dimension, consistent with our findings in this paper and contrary to many widely-held expectations; i.e., our notion of a resolvable transition from large to small polaron structure persists into the adiabatic limit.

Less firm information is available regarding the occurrence of critical self-trapping in higher dimensions; however, in dimensions $[D]$ dynamical mean field theory \cite{Ciuchi95,Ciuchi97}, for example, offers the estimates $\lambda_c[1] = 1/2$, $\lambda_c[2] \approx 0.8$, and $\lambda_c[\infty] \approx 0.844$, while our own estimates based on scaling arguments \cite{Romero99,Romero98f} suggest $\lambda_c [1] = 1/2$, $\lambda_c [2] \approx 0.8536$, $\lambda_c [3] \approx 0.9082$, and $\lambda_c [ \infty ] = 1$.
Both sets of estimates support the notion that at least the existence of the self-trapping transition and the scaling properties related to it are not seriously affected by changes in dimensionality.

Thus, adiabatic theory appears to be exceptional in suggesting a sharp distinction between polaron properties in one vs. higher dimensions.

\section{Conclusion}

In this paper, we have presented a large volume of data, comprehensive in scope and highly accurate, for the ground state energy $E_0$, kinetic energy $E_{kin}$, phonon energy $E_{ph}$, and electron-phonon interaction energy $E_{int}$ for the Holstein molecular crystal model of a single electron in one dimension.

We have broadly demonstrated the agreement between these results and the appropriate perturbation theories at weak and strong coupling, and have demonstrated the breakdown of both weak- and strong-coupling perturbation theory in the intermediate regime.
For particular cases we have demonstrated the detailed mutual consistency of our results with those of the density matrix renormalization group and with those of quantum Monte Carlo simulation from weak coupling to strong coupling.

Thus amply confirmed, we have analyzed the dependence of each of the principal energies upon the electron-phonon coupling constant to determine the self-trapping transition, defined relative to each measured property as the point at which that property experiences its most rapid change with respect to $g$.
The data thus collected cluster in a clear way, suggesting the definition of a single self-trapping line $g_{ST} = 1+\sqrt{J/\hbar\omega}$.
This line joins with the adiabatic critical point in the adiabatic limit, and separates large polaron structure from small polaron structure at all $J/\hbar\omega$.

Whether one chooses to consider the specific data appropriate to one physical property or the $g_{ST}$ characteristic of all, the self-trapping points here determined are consistent with features found in the particular kinetic energies, ground state energies, correlation functions, and effective masses determined independently by other high-quality methods.

These findings support the notion of the self-trapping transition as a smooth phenomenon at finite $J/\hbar\omega$ and $g$, sharpening steadily as one approaches the adiabatic limit.
As such, the self-trapping line associated with this transition is not unique, but is accurately and robustly located by $g_{ST}$.

\section*{acknowledgment}

This work was supported in part by the U.S. Department of Energy under Grant No.
 DE-FG03-86ER13606.
The authors gratefully acknowledge the cooperation of P. E. Kornilovitch and E. Jeckelmann for providing numerical values of data used in parts of this paper.

\bibliography{/home/bassi/dwb/Tex/Bibliography/theory,/home/bassi/dwb/Tex/Bibliography/books,/home/bassi/dwb/Tex/Bibliography/experiment,/home/bassi/dwb/Tex/Bibliography/temporary}

\begin{thebibliography}{10}

\bibitem{Migdal58}
A.~B. Migdal, Sov. Phys. JETP {\bf 34},  996  (1958).

\bibitem{Nakajima80}
S. Nakajima, Y. Toyozawa, and R. Abe, {\em The Physics of Elementary
  Excitations} (Springer-Verlag, Berlin, 1980).

\bibitem{Mahan93}
G.~D. Mahan, {\em Many-Particle Physics} (Plenum Press, New York, 1993).

\bibitem{Alexandrov95}
A.~S. Alexandrov and S.~N. Mott, {\em Polarons \& Bipolarons} (World
  Scientific, London, 1995).

\bibitem{Romero98d}
A.~H. Romero, D.~W. Brown, and K. Lindenberg, Phys. Lett. A {\bf XX},  to
  appear  (1999).

\bibitem{Romero98g}
A.~H. Romero, Ph.D. thesis, University of California, San Diego, 1998.

\bibitem{Romero99}
A.~H. Romero, D.~W. Brown, and K. Lindenberg, in preparation  (1999).

\bibitem{Tyablikov52}
S.~V. Tyablikov, Zh. Eksp. Teor. Fiz. {\bf 23},  381  (1952).

\bibitem{Lang63}
I.~G. Lang and Y.~A. Firsov, Sov. Phys. JETP {\bf 16},  1301  (1963).

\bibitem{Gogolin82}
A.~A. Gogolin, Phys. Stat. Sol. (b) {\bf 109},  95  (1982).

\bibitem{Marsiglio95}
F. Marsiglio, Physica C {\bf 244},  21  (1995).

\bibitem{Stephan96}
W. Stephan, Phys. Rev. B {\bf 54},  8981  (1996).

\bibitem{Capone97}
M. Capone, W. Stephan, and M. Grilli, Phys. Rev. B {\bf 56},  4484  (1997-II).

\bibitem{Brown97b}
D.~W. Brown, K. Lindenberg, and Y. Zhao, J. Chem. Phys. {\bf 107},  3179
  (1997).

\bibitem{Brown97a}
D.~W. Brown and K. Lindenberg, Physica D {\bf 113},  267  (1998).

\bibitem{Zhao97a}
Y. Zhao, D.~W. Brown, and K. Lindenberg, J. Chem. Phys. {\bf 106},  5622
  (1997).

\bibitem{Zhao97b}
Y. Zhao, D.~W. Brown, and K. Lindenberg, J. Chem. Phys. {\bf 107},  3159
  (1997).

\bibitem{Romero98a}
A.~H. Romero, D.~W. Brown, and K. Lindenberg, J. Chem. Phys. {\bf 109},  6540
  (1998).

\bibitem{Romero98e}
A.~H. Romero, D.~W. Brown, and K. Lindenberg, Phys. Rev. B {\bf XX},  to appear
   (1999).

\bibitem{Trugman99}
S.~A. Trugman and J. Bonca, J. Supercon. {\bf 12},  221  (1999).

\bibitem{Bonca98}
J. Bonca, S.~A. Trugman, and I. Batistic, cond-mat 9812252  (1998).

\bibitem{Wellein97a}
G. Wellein and H. Fehske, Phys. Rev. B {\bf 56},  4513  (1997).

\bibitem{deMello97}
E.~V.~L. de~Mello and J. Ranninger, Phys. Rev. B {\bf 55},  14872  (1997).

\bibitem{Alexandrov94a}
A.~S. Alexandrov, V.~V. Kabanov, and D.~E. Ray, Phys. Rev. B {\bf 49},  9915
  (1994).

\bibitem{Jeckelmann98a}
E. Jeckelmann and S.~R. White, Phys. Rev. B {\bf 57},  6376  (1998).

\bibitem{Jeckelmann98b}
E. Jeckelmann, private communication  (1998).

\bibitem{McKenzie96}
R.~H. McKenzie, C.~J. Hamer, and D.~W. Murray, Phys. Rev. B {\bf 53},  9676
  (1996).

\bibitem{DeRaedt83}
H.~D. Raedt and A. Lagendijk, Phys. Rev. B {\bf 27},  6097  (1983).

\bibitem{DeRaedt84}
H.~D. Raedt and A. Lagendijk, Phys. Rev. B {\bf 30},  1671  (1984).

\bibitem{Lagendijk85}
A. Lagendijk and H.~D. Raedt, Phys. Lett. A {\bf 108A},  91  (1985).

\bibitem{Kornilovitch98a}
P.~E. Kornilovitch, Phys. Rev. Lett. {\bf 81},  5382  (1998).

\bibitem{Alexandrov98a}
A.~S. Alexandrov and P.~E. Kornilovitch, Phys. Rev. Lett. {\bf 82},  807
  (1998).

\bibitem{Kornilovitch99}
P.~E. Kornilovitch, private communication  (1999).

\bibitem{Holstein59a}
T. Holstein, Ann. Phys. (N.Y.) {\bf 8},  325  (1959).

\bibitem{Holstein59b}
T. Holstein, Ann. Phys. (N.Y.) {\bf 8},  343  (1959).

\bibitem{Alexandrov98}
A. Alexandrov, V. Kabanov, E. Kudinov, and Y.~A. Firsov, cond-mat 9801117
  (1997).

\bibitem{Lowen88}
H. L\"owen, Phys. Rev. B {\bf 37},  8661  (1988).

\bibitem{Sumi73}
A. Sumi and Y. Toyozawa, J. Phys. Soc. Japan {\bf 35},  137  (1973).

\bibitem{Emin76}
D. Emin and T. Holstein, Phys. Rev. Lett. {\bf 36},  323  (1976).

\bibitem{Toyozawa80a}
Y. Toyozawa and Y. Shinozuka, J. Phys. Soc. Jap. {\bf 48},  472  (1980).

\bibitem{Kabanov93}
V.~V. Kabanov and O.~Y. Mashtakov, Phys. Rev. B {\bf 47},  6060  (1993).

\bibitem{Kabanov94}
V.~V. Kabanov and D.~K. Ray, Phys. Lett. A {\bf 186},  438  (1994).

\bibitem{Ciuchi95}
S. Ciuchi, F.~D. Pasquale, and D. Feinberg, Euro. Lett. {\bf 30},  151  (1995).

\bibitem{Ciuchi97}
S. Ciuchi, F. de~Pasquale, S. Fratini, and D. Feinberg, Phys. Rev. B {\bf 56},
  4494  (1997).

\bibitem{Romero98f}
A.~H. Romero, D.~W. Brown, and K. Lindenberg, MRS Symp. Proc. {\bf XX},  to
  appear  (1998).

\end{thebibliography}

\end{document}